\documentclass[%
aip,
jcp,%
amsmath,amssymb,
reprint,%
]{revtex4-2}

\usepackage[utf8]{inputenc}
\usepackage[T1]{fontenc}
\usepackage{graphicx}
\usepackage{dcolumn}
\usepackage{bm}
\usepackage{float}
\usepackage{lipsum}
\usepackage{siunitx}
\usepackage[colorlinks=true, citecolor=blue, linkcolor=black, urlcolor=blue, pdfborder={0 0 0}]{hyperref}
\usepackage{caption}
\usepackage{subcaption}

\usepackage[capitalise]{cleveref}

\newlength{\myfigwidth}

\newcommand{\ket}[1]{\left|#1\right\rangle}

\newcommand{\el}{\ell_{\textrm{loc}}}
\newcommand{\tl}{t_{\text{loc}}}

\newcommand{\Dinf}{D_{\infty}}
\newcommand{\VV}{\langle V^2\rangle}
\newcommand{\Wlgs}{W_{\textrm{LGS}}}

\begin{document}

\title{Temperature Dependence of Charge and Exciton Transport in One-Dimensional Systems Subject to Static and Dynamic Disorder}

\author{William Barford}
\email{william.barford@chem.ox.ac.uk}
\affiliation{Department of Chemistry, Physical and Theoretical Chemistry Laboratory, University of Oxford, Oxford, OX1 3QZ, United Kingdom}
\affiliation{Balliol College, University of Oxford, Oxford, OX1 3BJ, United Kingdom}

\date{\today}

\begin{abstract}
The  temperature-dependence of dynamical properties (e.g., the asymptotic diffusion coefficient and the sub-diffusive exponent) are calculated for  charges and excitons in one-dimensional systems subject to static and dynamic disorder.
These properties are  determined by three complementary methods. One approach is via the time-integration of the velocity autocorrelation function. The second  is via the mean-squared-displacement of thermal wavepackets subject to stochastic collapse via Lindblad jump operators. These two methods are applicable in the high-temperature regime, where the noise is temporally uncorrelated. In this regime the noise causes particle localization and the transport is diffusive. The third approach -- applicable in the low-temperature regime -- is  weak-coupling Redfield theory. Here, static disorder causes particle localization. When the dynamics is diffusive, the diffusion coefficient is a non-monotonic function of temperature, increasing with temperature in the low-temperature Environment Assisted Quantum Transport regime and decreasing with temperature in the high-temperature quantum-Zeno regime. For any temperature, static and dynamic disorder decreases the diffusion coefficient.
The dynamics is non-diffusive for thermal energies deep within the manifold of local-ground-states, where the sub-diffusive exponent decreases with increasing disorder and decreasing temperature.
\end{abstract}

\maketitle


\section{Introduction}\label{Se:1}

Charge and energy transport in low-dimensional  molecular materials, e.g., J-aggregates and $\pi$-conjugated polymers, is affected by a wide range of processes. Intrinsic static disorder causes a ballistically evolving wavepacket to become Anderson localized as a consequence of coherent superposition. Conversely, extrinsic dynamical disorder destroys coherences, causing diffusive propagation of the wavepacket. Sufficiently weak dephasing in disordered systems enhances diffusion\cite{Thouless1981,Logan1987,Wolynes1990}  - a phenomenon known as Environment Assisted Quantum Transport (EAQNT)\cite{Aspuru2009,Blach2025}; whereas strong dephasing suppresses diffusion\cite{Thouless1981,Logan1987,Wolynes1990}  - a phenomenon sometimes referred to as the quantum-Zeno (QZ) effect. In addition to static and dynamic disorder, polaronic effects caused by  electron-phonon coupling also affect transport properties. Understanding the mechanisms of charge and energy transport in molecular systems is theoretically challenging, but it is also necessary for predicting structure-function relationships, and hence optimizing the performance of molecular optoelectronic devices.

In the  high-temperature limit, defined by the temperature exceeding the  particle bandwidth, the role of static and dynamic disorder has been extensively investigated via the Haken-Strobl-Reineker (HSR) model\cite{Aspuru2009,Cao2013,Knoester2021}. The HSR model is a simplified description of system-bath interactions, in which a classical, Markovian bath causes  white-noise dynamical fluctuations of the particle site energies\cite{Haken1973,Reineker}.
The HSR stochastic quantum Liouville equation  with a white-noise spectrum causes system Hamiltonian eigenstate  populations to evolve to equal values. As a consequence, except from the temperature-dependence of the dephasing rate, the HSR-limit provides no physical description of the temperature dependence of dynamical properties for temperatures lower than the particle bandwidth.

In this paper we develop a theory of charge and energy transport as a function of temperature. The temperature of the system is maintained in three different ways. First,  this is done explicitly when using the velocity autocorrelation function method (see Section \ref{Se:2.2.1}), because the Boltzmann equilibrium density operator is used when computing the asymptotic diffusion coefficient, $\Dinf(T)$. Second,  when $\Dinf(T)$ is computed via the mean-squared-displacement  using the eigenstate thermalization hypothesis (ETH) (see Section \ref{Se:2.2.2}), this is again explicit for uniform systems, because the ETH creates a Boltzmann distribution of energy eigenstates. However,  this condition fails for disordered systems at low temperatures. Finally, when  low temperature transport properties are computed using the Redfield equation (see Section \ref{Se:2.3}), a constant temperature is maintained by the form of the interstate rates, and  mean energy fluctuations are minimized by the choice of the initial energy eigenstate.

Maintaining the system at a finite temperature gives a quantitatively different description of particle dynamics than is obtained from the `high-T' limit (i.e., the limit of equal energy eigenstate populations). In particular, the diffusion coefficient is reduced, the EAQNT to QZ crossover occurs at a higher temperature, and  the low temperature dynamics is sub-diffusive. This paper is focussed on the roles of temperature, and static and dynamic disorder. Thus, the important role of polaronic effects are neglected, but are described in various reviews\cite{Kohler2015,Troisi2021,Blumberger2022,Barford2022}.

The next section contains the theoretical background to this work. First, as a prototypical model of particle dynamics in one-dimensional systems,  we introduce the disordered Frenkel exciton Hamiltonian, and then describe the effect of static disorder on localizing the exciton eigenstates. Next, we describe the system-bath coupling, and introduce the quantum and classical bath autocorrelation functions that quantify the dynamical disorder. The two complementary  methods for computing the diffusion coefficient in the high-temperature limit -- where dephasing dominates -- namely via the velocity autocorrelation function and the ETH, are described in Section \ref{Se:2.2}. Section \ref{Se:2.3} describes weak-coupling Redfield theory, which is used to compute dynamical properties in the low-temperature limit where static disorder dominates. The results of all the methods are presented and discussed in Section \ref{Se:3}, while Section \ref{Se:4} concludes the paper.

\section{Theoretical Details}\label{Se:2}

\subsection{Model of Exciton Dynamics in Linear Molecular Systems}\label{Se:2.1}

We formulate the problem of charge and energy transport one-dimensional  systems in terms of Frenkel exciton dynamics in  molecular systems, e.g., J-aggregates or $\pi$-conjugated polymers. However, the analysis applies equally to triplet excitons and charges.

The total Hamiltonian is
\begin{equation}\label{Eq:1}
  \hat{H} = \hat{H}_S + \hat{H}_{SB} + \hat{H}_B,
\end{equation}
where $\hat{H}_S$,  $\hat{H}_{SB}$ and $\hat{H}_B$ are the system, system-bath and bath Hamiltonians, respectively.
The system Hamiltonian is
\begin{equation}\label{}
  \hat{H}_S = \sum_{n=1}^N  \epsilon_n |n\rangle \langle n| - J\sum_{n=1}^N \left(|n+1\rangle \langle n| + |n\rangle \langle n+1|\right),
\end{equation}
where the ket $|n\rangle$ represents an exciton on monomer $n$, denoting a `site' and $N$ is the number of sites.
$\epsilon$  and $J$ are the onsite potential and  nearest-neighbor transfer integral, respectively. $J$ is taken to be uniform and positive, while $\epsilon_n$ encodes the static disorder by being an uncorrelated Gaussian random variable of variance $\sigma^2$ with a zero-mean value.

The eigenstates of $\hat{H}_S$ are denoted as
\begin{equation}\label{Eq:30}
  |a\rangle = \sum_n \psi_{na} |n\rangle
\end{equation}
with eigenvalues $E_a$. All eigenstates of a disordered  one-dimensional system are spatially Anderson localized\cite{Anderson1958,Mott1961}. However, as shown by Malyshev and Malyshev\cite{Malyshev2001a,Malyshev2001b}, the low-energy spectrum is comprised of super-localized states, that they named local ground states (LGS). LGS (defined by the condition that $|\sum_n \psi_{na}|\psi_{na}|| > 0.95$) are essentially nodeless, space filling and spatially nonoverlapping. Their size scales with disorder as $\sigma^{-2/3}$, and thus (via exchange narrowing\cite{Knapp1984}) their band width scales $W_{\textrm{LGS}} \sim J(\sigma/J)^{4/3}$. Higher energy quasi-extended states are nodeful and denoted as quasi-extended states (QES).
The energy density of states for LGS and for all states near to the band edge are illustrated in Fig.\ 1.
\begin{figure}[tb]
\includegraphics[width=0.9\linewidth]{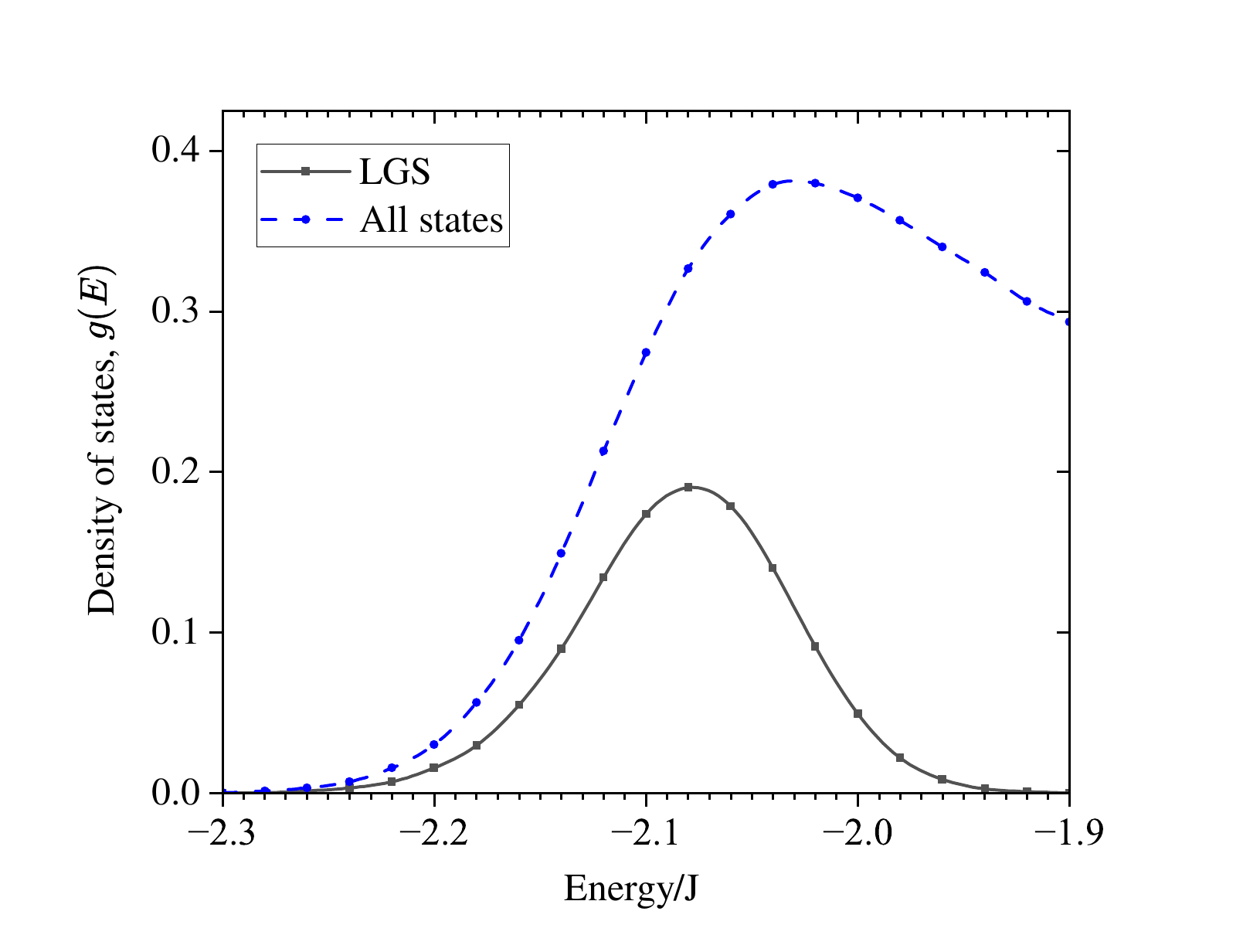}\label{Fi:4}
\caption{Density of states, $g(E)$ (where $\int g(E) \textrm{d}E = 1$), for LGS and for all states near to the band edge (at $E = -2J$ when $\sigma =0$).
The width of the LGS density of states is $W_{\textrm{LGS}} \sim  J(\sigma/J)^{4/3}$.
For these results the static on-site energy disorder is $\sigma/J = 0.2$.}
\end{figure}

The system-bath Hamiltonian is
\begin{eqnarray}\label{Eq:4}
\hat{H}_{SB} = \sum_{n} \delta \epsilon_n(t)|n\rangle\langle n|,
\end{eqnarray}
where $\delta \epsilon_n(t)$ is the dynamical fluctuation of the on-site potential energy.
For linear system-bath coupling\cite{Nitzan}
\begin{equation}\label{}
  \delta \epsilon_n(t) = \sum_j c_{nj} u_j(t),
\end{equation}
where $u_j$ are the mass-weighted oscillator displacements with angular frequency $\omega_j$ and
$c_{j}$ are the weightings of each normal mode with an associated spectral function
\begin{equation}\label{}
  J(\omega) =  \frac{\pi}{2} \sum_j \frac{c_j^2}{\omega_j} \delta(\omega-\omega_j).
\end{equation}
Finally, the bath Hamiltonian of harmonic oscillators is\cite{Nitzan}
\begin{eqnarray}\label{Eq:47}
\hat{H}_{B} = \frac{1}{2}\sum_{j} \left(\dot{u}_j^2 + \omega_j^2 u_j^2\right).
\end{eqnarray}

Assuming spatially  uncorrelated site energy fluctuations, the bath autocorrelation function is
\begin{equation}\label{}
  C(t) =  \langle \delta \epsilon_m(t)  \delta \epsilon_n(0) \rangle \delta_{mn}.
\end{equation}
For a quantum harmonic bath  and linear system-bath coupling\cite{Nitzan}
\begin{widetext}
\begin{eqnarray}\label{Eq:90}
 C_{Q}(t) = \frac{\hbar}{\pi} \int_0^{\infty}  J(\omega)\left[(n(\omega)+1)\exp(\textrm{-i}\omega t) + n(\omega)\exp(\textrm{i}\omega t)\right] \textrm{d} \omega,
\end{eqnarray}
\end{widetext}
where $n(\omega) = (\exp(\hbar \omega/k_B T) -1)^{-1}$ is the Bose distribution function.
For a classical  bath, defined by $\hbar \omega \ll k_B T$, Eq.\ (\ref{Eq:90}) becomes,
\begin{eqnarray}\label{Eq:100}
 C_{C}(t) = \frac{2k_B T}{\pi} \int_0^{\infty} \frac{J(\omega)}{\omega} \cos(\omega t) \textrm{d} \omega.
\end{eqnarray}
The spectral function for an Ohmic bath with a high-frequency cut-off, $\omega_0$, is defined as
\begin{equation}\label{}
  J(\omega) =  \left(\frac{\pi \lambda \omega}{\omega_0} \right) \exp(-\omega/\omega_0),
\end{equation}
where $\lambda  = \sum_j {c_j^2}/2{\omega_j^2}$ is the reorganization energy of the bath modes.

In this paper we investigate the role of dynamical fluctuations in the white-noise limit, defined by $\omega \ll \omega_0$. In this case
\begin{equation}\label{Eq:1300a}
  J(\omega) \rightarrow \pi \lambda\omega/\omega_0
\end{equation}
and thus $J(\omega) \sim \omega I \sim \omega$, where $I$ is the frequency-independent power spectrum. This implies
that $\hat{H}_{SB}(t)$ induces transitions between all pairs of eigenstates.

In the white-noise limit using the spectral function defined in Eq.\ (\ref{Eq:1300a}), the classical bath auotcorrelation function, Eq.\ (\ref{Eq:100}), becomes uncorrelated in time, i.e.,
\begin{eqnarray}\label{}
 C_{C}(t) = \Gamma \hbar^2 \delta(t),
\end{eqnarray}
where\cite{Barford2024}
\begin{equation}\label{eq:15}
  \Gamma =  \frac{2\pi k_B T  \lambda}{\hbar^2 \omega_0}
\end{equation}
is the dephasing rate.
For later, it is useful to define $\Gamma_0$ such that $\Gamma = \Gamma_0 \times T$, i.e.,
\begin{equation}\label{Eq:16}
  \Gamma_0 =  \frac{2\pi k_B  \lambda}{\hbar^2 \omega_0}.
\end{equation}

\subsection{Determining the Diffusion Coefficient in the High-Temperature Limit:  $T > J$}\label{Se:2.2}

This section describes the methods by which  the asymptotic diffusion coefficient is calculated for a classical bath in the white-noise limit. First, via the velocity autocorrelation function and second, via thermalized eigenstates and Lindblad jump operators.

In a diffusive process the diffusion coefficient is defined as
\begin{equation}\label{}
D(t) = \frac{1}{2}\frac{\textrm{d}\textrm{MSD}(t)}{\textrm{d}t},
\end{equation}
where the mean-squared-displacement (MSD) is
\begin{equation}\label{}
  \textrm{MSD}(t) = \int_0^t\int_0^t V(t')V(t'')\textrm{d}t'\textrm{d}t''
\end{equation}
and $V(t)$ is the particle velocity.

\subsubsection{Using the Velocity Autocorrelation Function}\label{Se:2.2.1}

Defining the velocity autocorrelation function as $C(t',t'') = V(t')V(t'')$, then if a system satisfies  \textit{stationarity}, i.e., if  $C(t+t',t')$ is independent of $t'$, one can show that
\begin{equation}\label{}
 \frac{\textrm{d}\textrm{MSD}(t)}{\textrm{d}t} = 2\int_0^t C_V(t')\textrm{d}t',
\end{equation}
and thus the thermal diffusion coefficient is defined as
\begin{equation}\label{Eq:B1}
D(t,T)=
\int_0^t \textrm{Tr}\{ \hat{\rho}(t',T)  \hat{C}_V(t')\}\textrm{d}t'.
\end{equation}

The asymptotic diffusion coefficient is therefore
\begin{equation}\label{Eq:8}
D_{\infty}(T)=
\int_0^{\infty} \textrm{Tr}\{ \hat{\rho}^{\textrm{B}}(T)  \hat{C}_V(t)\}\textrm{d}t,
\end{equation}
where
\begin{equation}\label{}
 \hat{\rho}^{\textrm{B}}(T) = \frac{\exp(-\beta \hat{H}_S)}{\textrm{Tr}\{\exp(-\beta \hat{H}_S)\}}
\end{equation}
is the  equilibrium density operator and $\beta = (k_\textrm{B}T )^{-1}$.

Setting $d = \hbar = 1$, the velocity operator on a lattice  is
\begin{equation}\label{}
 \hat{V} = \textrm{i} J \sum_n \left( |n\rangle\langle n + 1 | - |n+1 \rangle\langle n  | \right).
\end{equation}
In the presence of temporally uncorrelated dissipation the Heisenberg representation of $\hat{V}$ satisfies the equation of motion (with $\hbar = 1$),\cite{Breuer,Knoester2016}
\begin{equation}\label{}
  \frac {\textrm{d}\hat{V}}{\textrm{d}t} = \textrm{i}[\hat{H}_S, \hat{V}] + \frac{\Gamma}{2} \sum_n \left[ [\hat{A}_n,\hat{V}],\hat{A}_n\right],
\end{equation}
where the second term on the right-hand-side is the Lindblad dissipator and $\hat{A}_n$  is the Lindblad jump operator. Expanding the commutators and using $\hat{A}_n = |n\rangle\langle n |$ for white-noise gives,
\begin{equation}\label{}
  \frac {\textrm{d}\hat{V}}{\textrm{d}t} = \textrm{i}\left(\hat{H}_S\hat{V}- \hat{V}\hat{H}_S\right) -\Gamma\hat{V}.
\end{equation}
Thus,
\begin{equation}\label{Eq:310}
  \hat{V}(t) = \exp(\textrm{i} \hat{H}_S t) \hat{V}(0) \exp(-\textrm{i}\hat{H}_S t) \exp(-\Gamma t)
\end{equation}
and the velocity autocorrelation function becomes\cite{Knoester2016,Knoester2021,Barford2024}
\begin{equation}\label{Eq:260}
  \hat{C}_V(t) = \exp(\textrm{i}\hat{H}_S t) \hat{V}(0)  \exp(-\textrm{i}\hat{H}_S t))\hat{V}(0) \exp(-\Gamma t).
\end{equation}

Using  Eq.\ (\ref{Eq:260}), Eq.\ (\ref{Eq:8}) now becomes
\begin{widetext}
\begin{equation}\label{Eq:10}
D_{\infty}(T)=
\int_0^{\infty} \textrm{Tr}\{ \hat{\rho}^{\textrm{B}}(T) \exp(\textrm{i}\hat{H}_0 t) \hat{V}(0)  \exp(-\textrm{i}\hat{H}_0t))\hat{V}(0) \exp(-\Gamma t)\}\textrm{d}t.
\end{equation}
\end{widetext}
Inserting the resolution of the identity, $\sum_a |a\rangle \langle a| \equiv \hat{1}$, between pairs of operators and performing the trace over the eigenstates of $\hat{H}_S$, Eq.\ (\ref{Eq:10}) is
\begin{widetext}
\begin{equation}\label{Eq:11}
D_{\infty}(T)=
\int_0^{\infty} \sum_{abc} {\rho}_{ab}^{\textrm{B}}(T)  {V}_{bc} {V}_{ca} \exp[(\textrm{i}\Delta E_{bc}-\Gamma) t]  \textrm{d}t,
\end{equation}
\end{widetext}
where $\Delta E_{bc}  = (E_b-E_c)$.
Finally, computing the time-integral gives the following result for the temperature-dependent asymptotic diffusion coefficient,
\begin{equation}\label{Eq:12}
D_{\infty}(T) = \Gamma  \sum_{abc}\frac{{\rho}_{ab}^{\textrm{B}}(T)  {V}_{bc} {V}_{ca} }{\Gamma^2+\Delta E_{bc}^2}.
\end{equation}

\emph{(a) Translationally invariant systems}

In a translationally invariant system $[\hat{H}_S,\hat{V}]=0$ and thus $\hat{V}|a\rangle = V_{aa} |a\rangle$, where $V_{aa} = 2J\sin k_a$, $k_a = 2\pi a/N$, and $1 \le a \le N$. In this case Eq.\ (\ref{Eq:12}) simply becomes
\begin{eqnarray}\label{Eq:13}
D_{\infty}(T)= && \frac{1}{\Gamma}  \sum_{a}{\rho}_{aa}^{\textrm{B}}(T) | {V}_{aa}|^2 \nonumber\\
             =  && \langle V^2\rangle_{\textrm{th}}/\Gamma,
\end{eqnarray}
where ${\rho}_{aa}^{\textrm{B}}(T) = P_a^{\textrm{B}}(T)$ is the Boltzmann probability. Thus, $\langle V^2\rangle_{\textrm{th}}$ is the  thermal average of the  expectation value of the mean-squared speed.

In the high temperature limit, i.e., $T \gg J$, all the energy eigenstates are thermally accessible so that $\langle V^2\rangle_{\textrm{th}} = 2J^2$ and $D_{\infty}(T) = 2J^2/\Gamma(T)$. This is the HSR result\cite{Reineker}. Conversely for $J/N \ll T \ll J$ (i.e., the spectrum is the quasicontinuous energy of a free particle) the classical limit applies so that $\langle V^2\rangle_{\textrm{th}} = 2J T$ and $D_{\infty}(T) = 2JT/\Gamma(T)$. Replacing $\Gamma(T)$ by $\Gamma_0 \times T$ gives the high temperature prediction that for translationally invariant systems $D_{\infty}(T) = 2J^2/(\Gamma_0 \times T)$, while at low temperatures $D_{\infty}(T) = 2J/\Gamma_0$, i.e., independent of temperature.

\emph{(b) Disordered systems}

For disordered systems $[\hat{H}_S,\hat{V}]\ne 0$ and so Eq.\ (\ref{Eq:12}) does not contract into a sum over the diagonal matrix elements. Indeed, the thermalized eigenstate coherences, ${\rho}_{ab}^{\textrm{eq}}(T)$, appear to contribute to $D_{\infty}(T)$.  In the HSR model the eigenstate coherences satisfy  the quantum Liouville equation\cite{Reineker}
\begin{eqnarray}\label{Eq:17}
 \frac{ \textrm{d}{\rho}_{ab}}{\textrm{d} t} &&=
 -   (\textrm{i}\Delta E_{ab} +\Gamma_{ab}){\rho}_{ab} + \sum_{cd\ne ab} R_{ab,cd} {\rho}_{cd}.
\end{eqnarray}
In a disordered system the second term on the RHS of Eq.\ (\ref{Eq:17}) couples coherences to populations. It also implies that in general eigenstate coherences do not vanish in the asymptotic limit. To make progress we now invoke the secular approximation\cite{May} and set $R_{ab,cd} = 0$, thus decoupling coherences and populations. Crucially, this approximation  implies that $\rho_{ab}(t) = \rho_{ab}(0)\exp[-(\textrm{i}\Delta E_{ab} +\Gamma_{ab})t]$ and hence eigenstate coherences now vanish in the asymptotic limit.
In addition, the secular approximation ensures that the populations equal their Boltzmann values in the asymptotic limit.
With this approximation, we return to Eq.\  (\ref{Eq:12}) and set $\rho_{ab}^{\textrm{B}}(T) = \rho_{ab}^{\textrm{B}}(T)\delta_{ab} \equiv P_a^{\textrm{B}}(T)$, so that
\begin{equation}\label{Eq:15}
D_{\infty}(T)= \Gamma(T)  \sum_{ab}\frac{ P_a^{\textrm{B}}(T) | {V}_{ab}|^2 }{\Gamma(T)^2+\Delta E_{ab}^2}.
\end{equation}
This form of $D_{\infty}(T)$ was used in ref\cite{Knoester2016,Knoester2021} in the high-temperature limit where $P_a^{\textrm{B}}(T) \rightarrow 1/N$.


\subsubsection{Using Thermalized Eigenstates and Lindblad Jump Operators}\label{Se:2.2.2}

An alternative method to determine the diffusion coefficient  is via the definition
\begin{equation}\label{}
D_{\infty}(T) = \frac{\textrm{MSD}(t)}{2t}|_{\textrm{limit }t\rightarrow \infty}.
\end{equation}
$\textrm{MSD}(t)$ is the mean-squared-displacement of a particle in a time $t$, which can be defined by the spread of a particle's wavefunction. A thermal diffusion coefficient implies a thermalized wavefunction. Here we adopt the eigenstate thermalization hypothesis\cite{ETH} (ETH) which states that ensemble averages of expectation values of an observable $O$ using a state vector
\begin{equation}\label{Eq:20}
  |\Psi_{\beta}\rangle  = \exp(-\beta \hat{H}_S/2) |\Psi\rangle
\end{equation}
is equivalent to the canonical ensemble average, $\langle O  \rangle = \textrm{Tr}\{  \hat{\rho}^{\textrm{B}}(T)  \hat{O}\}$.

A difficulty with evolving wavefunctions with the Hamiltonian $\hat{H} = \hat{H}_S + \hat{H}_{SB}$, however, is that the  time-dependence of the Hamiltonian arising from white-noise fluctuations (i.e., $\hat{H}_{SB}(t)$) causes any wavefunction to evolve to its high-temperature limit, i.e., all the energy eigenstates become equally populated. This difficulty can be avoided by evolving the state vector via the time-independent part (i.e., $\hat{H}_S$) and imposing  white-noise via Lindblad jump operators. In particular, the role of the Lindblad dissipator  in the stochastic quantum Liouville equation\cite{Breuer}
can be simulated by an ensemble of quantum trajectories  if each trajectory undergoes a non-unitary evolution via an effective non-hermitian Hamiltonian, defined by\cite{Daley2014}
\begin{eqnarray}
 \nonumber
  \hat{H}_S \rightarrow \hat{H}_{\textrm{eff}}  = && \hat{H}_S  - \frac{\textrm{i} \Gamma}{2} \sum_m \hat{A}_m^{\dagger}\hat{A}_m,
\end{eqnarray}
where $\hat{A}_m$ is the Lindblad jump operator.
For diagonal white-noise, $\hat{A}_m = |m\rangle\langle m|$.\cite{Aspuru2009}
The simulation of a quantum trajectory then proceeds as follows:
\begin{enumerate}
\item{ Given $|\Psi(t)\rangle$ at a time $t$, compute
\begin{equation}\label{}
  |\Psi_{\textrm{trial}}\rangle = \exp(-\textrm{i}\hat{H}_{\textrm{eff}} \delta t) |\Psi(t)\rangle,
\end{equation}
where $\delta t$ is the time step.}
\item{Determine $\delta p$, defined via $\langle \Psi_{\textrm{trial}}|\Psi_{\textrm{trial}}\rangle = 1 - \delta p$.}
\item{Then,\\ (i) with a probability $(1-\delta p)$ define
\begin{equation}\label{}
  |\Psi(t+\delta t)\rangle = |\Psi_{\textrm{trial}}\rangle/(1-\delta p)^{1/2}.
\end{equation}
In this case the wavefunction evolves according to $\hat{H}_{\textrm{eff}}$.\\
 Or, \\
(ii) with a probability $\delta p$ define
\begin{eqnarray}\label{Eq:A14}
  |\Psi(t+\delta t)\rangle &&= \frac{\hat{A}_m |\Psi(t)\rangle}{\langle \Psi(t| \hat{A}_m^{\dagger}\hat{A}_m |\Psi(t)\rangle^{1/2}}.
\end{eqnarray}
Using  the definition that $\hat{A}_m = |m\rangle\langle m|$, Eq.\ (\ref{Eq:A14}) implies that $|\Psi(t+\delta t)\rangle    = |m\rangle$. Thus, in this case the wavefunction collapses onto site $m$.
}
\item{If a quantum jump occurs in 3(ii), the site $m$ is chosen with a probability $P_m = |\Psi_m(t)|^2$.}
\end{enumerate}

Thus, the eigenstate  thermalization hypothesis is that an ensemble of quantum states $|\Psi_{\beta}\rangle$ evolved subject to the protocol outlined above reproduces thermal averages of observables of a system subject to white-noise. As we now show, this hypothesis appears to be rigorously accurate for translationally invariant systems, but breaks down for disordered systems as $T \rightarrow 0$.

To illustrate this point, let us consider a thermal state constructed from a delta-function source, i.e., $|\Psi\rangle = |m\rangle$. Using Eq.\ (\ref{Eq:30}) it is easy to show that $|\Psi_{\beta}\rangle \equiv  \exp(-\beta \hat{H}_S/2) |m\rangle = \sum_n c_n |n\rangle$ where
\begin{equation}\label{Eq:26}
  c_n = \sum_a \psi_{ma}^*\psi_{na} \exp(-\beta E_a/2).
\end{equation}

For a translationally invariant system, the transformation matrix elements are the Bloch factors, i.e.,
\begin{equation}\label{Eq:130}
\psi_{na}  = (\psi^{-1})^*_{an} = \frac{1}{\sqrt{N}} \exp(-\textrm{i}k_a n).
\end{equation}
Thus, $c_n$ is the Bloch-Fourier transform in k-space of $\tilde{c}_k = \exp(-\beta E_a/2)$.
To understand its physical significance, we return to dimensionful variables. Then,
in the continuum limit $E_a \rightarrow J ((k_a d)^2 -2)$, so that $\tilde{c}_k$ is a Gaussian function whose probability distribution  has a mean-squared-width $\Delta k^2 = 1/2\beta J d^2$. Finally, replacing $J$ by $\hbar^2/2md^2$, we find that $m\Delta V^2/2= \hbar^2 \Delta k^2/2m = k_B T/2$, in accordance with the principle of equipartition.

We can also investigate whether the energy expectation value equals the canonical ensemble result. Using Eq.\ (\ref{Eq:20}),
\begin{eqnarray}\label{Eq:28}
  \langle E_{\Psi} \rangle =  && \frac{\langle  \Psi_{\beta}| \hat{H}_0 |\Psi_{\beta}\rangle}{\langle  \Psi_{\beta}|\Psi_{\beta}\rangle}\nonumber \\
=   && \frac{\sum_a |\psi_{ma}|^2 E_a \exp(-\beta E_a)}{\sum_a |\psi_{ma}|^2 \exp(-\beta E_a)}
\end{eqnarray}
Evidently, this expression only identically equals the Boltzmann expression for Bloch states, when $|\psi_{ma}|^2  = 1/N$.

As described in Section \ref{Se:2.1} and illustrated in Fig.\ 1, the low-energy spectrum is comprised of local ground states (LGS), whose band width $W_{\textrm{LGS}} \sim J(\sigma/J)^{4/3}$.
As we now show, for $T <  W_{\textrm{LGS}}$ the ETH fails.
We rewrite the thermal wavefunction, $c_n$ (Eq.\ (\ref{Eq:26})), as
\begin{equation}\label{}
  \Psi_{\beta}^m(n) = N \sum_a \psi_{ma}\psi_{na} \exp(-\beta \Delta E_a/2),
\end{equation}
where $m$ represents the point source,  $N$ is a normalization factor, $\psi_{na}$  is the eigenstate wavefunction and $\Delta E_a$ is the excitation energy  of $|a\rangle$ with respect to the lowest-energy eigenstate. For $\beta  W_{\textrm{LGS}} \gg 1$ only LGSs can contribute to the sum. However, since LGS are exponentially localized with only small overlap with next nearest neighbors, the summand will vanish for all states except that for which $\psi_{ma}$ is not negligible. (Higher-lying quasiextended states which do have a non-zero amplitude for $n=m$ are projected out of the sum by the Boltzmann factor.)

This point is illustrated in Fig.\ 2, which shows  the maximum value of the projection of $|\Psi_{\beta}\rangle$ onto a LGS ($|\psi_{LGS}\rangle$) for one realization of the disorder, which is then averaged over 100 realizations of  the disorder. As $T \rightarrow 0$ this projection approaches unity, i.e., only one eigenstate contributes to $|\Psi_{\beta}\rangle$. However, in general  this eigenstate will not be the global ground state; it is only the `ground state' at $m=n$. Consequently, $\langle E_{\Psi} \rangle$ does not equal the canonical ensemble value, $\langle E_{B} \rangle$, as shown in the inset of Fig.\ 2.

\begin{figure}[tb]\label{Fi:1}
\includegraphics[width=0.8\linewidth]{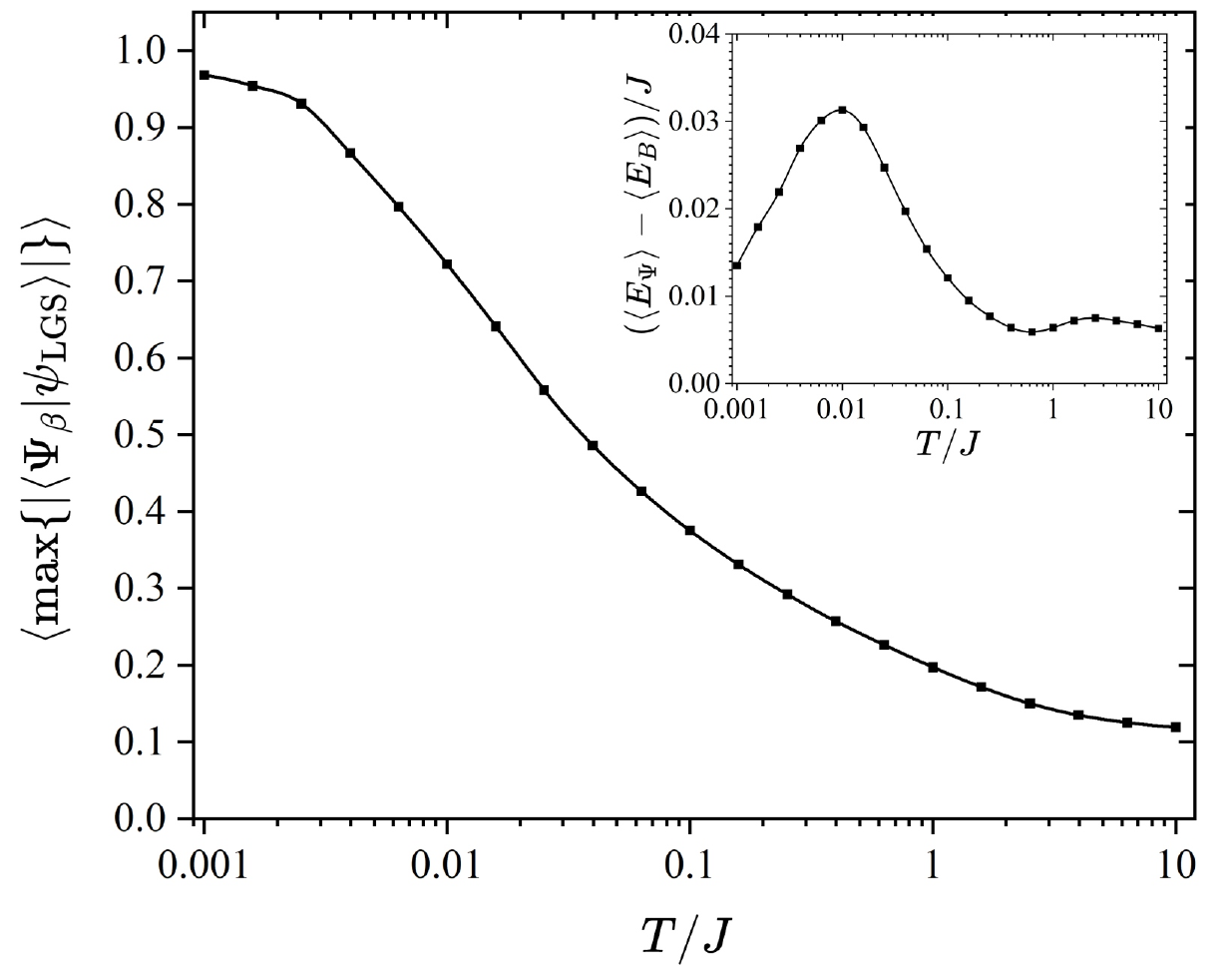}
\caption{The ensemble-averaged maximum projection of $|\Psi_{\beta}\rangle$ onto  a LGS, $|\psi_{LGS}\rangle$, as a function of temperature ($k_B = 1$). $\sigma/J = 0.1$, chains of 200 sites and 100 realizations of the disorder. The inset shows the ensemble-averaged deviation of $\langle E_{\Psi} \rangle$ (Eq.\ \ref{Eq:28}) from the Boltzmann average, $\langle E_{\textrm{B}} \rangle = \sum_a P_a^{\textrm{B}} E_a$.
}
\end{figure}

In the usual Linblad jump operator formalism the wavefunction collapses to a $\delta$-function at site $m$, whereupon its time evolution continues.  As shown in ref\cite{Barford2024} in the $T\rightarrow \infty$ limit, by defining the  MSD as the sum of the squares of the jump distances, this procedure reproduces the HSR prediction for $D_{\infty}$. For thermalized eigenstates, however, this protocol must be modified because the new thermal wavepacket after each collapse has an initial mean-squared size, $\Delta n(0)^2 \propto T^{-1}$. We thus define the MSD between jumps as the dispersion of the wavepacket, i.e., at the $i$th jump,  $\textrm{MSD}_i = \Delta n(\tau_i)^2 - \Delta n(0)^2$, where $\tau_i$ is the time interval of the $i$th jump. The total MSD in a time $t = \sum_i \tau_i$ is then defined as $\textrm{MSD} = \sum_i \textrm{MSD}_i$. To avoid boundary effects, after each jump  $m$ is set to $N/2$, so that the wavefunction is centered in the middle of the chain.
As shown in Section \ref{Se:3.1}, in the $T \gg J$ limit this new protocol also reproduces the HSR prediction.  For disordered systems a new realization of the disorder is generated after each jump and the Hamiltonian is again diagonalized.

The simulations described in Section \ref{Se:3.1} were performed by numerically solving the TDSE using  the \textit{Short Iterative Lanczos Propagator} method\cite{Park1986,Tannor,Barford2024}.

\subsection{Determining the Transport Coefficients in the Low-Temperature Limit: $T < J$}\label{Se:2.3}

For $T < J$ static disorder - causing particle localization - plays a key role in the particle dynamics. As shown in  Section \ref{Se:3.2}, transport via the localized and quasi-extended exciton states is sub-diffusive.
In addition, when $T \ll J$, the condition for a classical bath, i.e., $T \gg \omega_{ab}$, is no longer generally applicable.

Under these circumstances, exciton dynamics is more appropriated modeled via the Redfield master equation, where the rates are determined assuming a quantum bath autocorrelation function. Again, making the secular approximation, the Redfield equation reads,
\begin{equation}\label{Eq:48}
  \frac{\textrm{d}P_a(t)}{\textrm{d}t} = - \sum_{b \ne a} \left( k_{ab}P_a(t) - k_{ba} P_b(t) \right).
\end{equation}
According to weak-coupling Redfield theory\cite{Nitzan}, the rates are
\begin{equation}\label{}
  k_{ab} = \frac{1}{\hbar^2} S_{ab} \int_0^{\infty}  C(t) \exp(\textrm{i}\omega_{ab}t)\textrm{d}t
\end{equation}
where $C(t)$ is the bath autocorrelation function and
\begin{eqnarray}
   S_{ab} &=& \sum_m |\psi_{ma}|^2 |\psi_{mb}
   |^2
\end{eqnarray}
is the density overlap function.
Thus, inserting the quantum autocorrelation function, Eq.\ (\ref{Eq:90}), the rates (defined for $E_a > E_b$) are
\begin{equation}\label{Eq:190}
  k_{ab} = \frac{2}{\hbar} J(\omega_{ab}) \left[n(\omega_{ab}) + 1 \right] S_{ab}
\end{equation}
and
\begin{equation}\label{Eq:200}
  k_{ba} = \frac{2}{\hbar} J(\omega_{ab}) n(\omega_{ab}) S_{ab}.
\end{equation}

The master equation Eq.\ (\ref{Eq:48}) and the rates given in Eq.\ (\ref{Eq:190}) and Eq.\ (\ref{Eq:200}) (albeit with a Debye spectral density)  have also been used by Bednarz \emph{et al.}\cite{Bednarz2002,Bednarz2003,Bednarz2004} to model  spectroscopy  in disordered one-dimensional systems, Vlaming \emph{et al.}\cite{Vlaming2013} to model sub-diffusive dynamics and Carta \emph{et al.}\cite{Hildner2024} to model exciton dynamics in nanofibres.

Using Eq.\ (\ref{Eq:1300a}) for the spectral density in the white-noise limit and noting that  $n(\omega_{ab}) \rightarrow k_B T/\hbar \omega_{ab}$ for $ k_B T \gg\hbar \omega_{ab}$, Eq.\ (\ref{Eq:190}) and Eq.\ (\ref{Eq:200}) become
\begin{equation}\label{}
  k_{ab} =  k_{ba}  = \Gamma(T) S_{ab},
\end{equation}
which is the HSR limit and reproduces the high-temperature form of the dephasing rate used in Section \ref{Se:2.2}.
\footnote{Note that $\Gamma_0$ is the only free parameter, because $2J(\omega_{ab})/\hbar = \hbar \Gamma_0 \omega_{ab}/k_B$.}

Equation (\ref{Eq:48}) is solved by casting it into the form
\begin{equation}\label{Eq:48b}
  \frac{\textrm{d}P_a(t)}{\textrm{d}t} = - \sum_{b} K_{ab} P_b(t),
\end{equation}
where
\begin{equation}\label{}
 K_{ab}  = k_{ba} - \sum_{b'} k_{ab'}\delta_{ab}.
\end{equation}
The solution of Eq.\ (\ref{Eq:48b}) from linear alegbra is
\begin{equation}\label{}
  P_a(t) = \sum_{bc} Z_{ab} \exp (\lambda_b t) Z_{bc}^{-1} P_c(0),
\end{equation}
where $\textbf{Z}$ is the matrix whose columns are the eigenvectors of $\textbf{K}$, $\{ \lambda \}$ are the corresponding eigenvalues and $P_c(0)$ is an initial condition.

The rates given by Eq.\ (\ref{Eq:190}) and Eq.\ (\ref{Eq:200}) ensure that the evolution of the populations, $P_a$, via Eq.\ (\ref{Eq:48}) occurs at a definite temperature. Moreover, they satisfy the detailed balance condition that
\begin{equation}\label{}
  \frac{k_{ab}}{k_{ba}} = \exp((E_a-E_b)/k_B T),
\end{equation}
meaning that the system evolves to thermal equilibrium determined by  the Boltzmann distribution, $P_a^{\textrm{B}}(T)$.

When choosing initial conditions it is tempting to select a thermal eigenstate $|\Psi_{\beta}\rangle \equiv  \exp(-\beta \hat{H}_S/2) |m\rangle$, where $|m\rangle$ is chosen at random. However, as shown in Section \ref{Se:2.2.2} and Fig.\ 2, for temperatures $T < \Wlgs$ such a projection predominately selects a single LGS that spatially spans the site $m$ and whose energy will be distributed randomly within the band of LGS. This means that during the evolution towards thermal equilibrium the mean energy will fluctuate by an amount $\sim \Wlgs$, which is larger than $T$ for $T < \Wlgs$.
So instead, for a particular realization of the disorder, the initial state, $\ket{i}$, is chosen whose energy lies closest to the thermal average, $\langle E_{\textrm{B}} \rangle = \sum_a P_a^{\textrm{B}}(T) E_a$  This ensures that during the evolution the mean energy will fluctuate by an amount smaller than $T$.
Therefore, we set $P_c(0) = \delta_{ci}$ and thus
\begin{equation}\label{Eq:50}
  P_a(t) = \sum_{b} Z_{ab} \exp (\lambda_b t) Z_{bi}^{-1}.
\end{equation}

The  matrix $\textbf{K}$ is diagonalized via the LAPACK routine  \textsf{DGEEV}, and the matrix $\textbf{S}$ is inverted via the LAPACK routines  \textsf{DGETRF} and \textsf{DGETRI}.
In practice, for long chains, high disorder and low temperatures, the diagonalization of $\textbf{K}$ is sometimes subject to numerical inaccuracies. The accuracy of the diagonalization is checked by ensuring that the asymptotic solution of Eq.\ (\ref{Eq:50}), i.e., $P_a(t\rightarrow \infty) = Z_{ab_0}Z_{b_0i}^{-1}$, where $b_0$ is the eigenvector of  $\textbf{K}$ with the vanishing eigenvalue, equals the Boltzmann distribution, $P_a^{\textrm{B}}$, to an accuracy of better than $10^{-6}$.

\section{Results}\label{Se:3}

Throughout this paper, unless explicitly stated otherwise, $J$ sets the energy scale, while $\hbar$, $k_B$ and the lattice spacing, $d$, are set to unity. Thus $T$, $\Gamma$,  $D$ and $V$ (the particle speed) are all in units of $J$.
For a classical, harmonic bath with  linear system-bath coupling  $\Gamma(T) = \Gamma_0 \times T$, where in this paper $\Gamma_0$ is a parameter (defined in Eq.\ (\ref{Eq:16})). As both $\Gamma$ and $T$ are defined in units of $J$ (since $\hbar = k_B =1$), $\Gamma_0$ is  dimensionless.

Since the particle's dynamical properties as a function of temperature are partly controlled by its mean thermal energy,  $\langle E_{\textrm{B}} \rangle$, we display its ensemble-average  for different disorder values in Fig.\ 3. For $T  \gg J$  all energy eigenstates become thermally accessible, disorder becomes irrelevant and $\langle E_{\textrm{B}} \rangle \rightarrow 0$. For $T \ll J$, on the other and, the particle resides predominately in the manifold of LGS (as shown in Fig.\ 1). In both of these limits the heat capacity $ C = \textrm{d}\langle E_{\textrm{B}} \rangle/\textrm{d}T $ vanishes, whereas in the regime $\Wlgs  \lesssim T \lesssim J$ the particle is `classical' and therefore $C = 0.5$ (in units of $k_\textrm{B} = 1$).

The inset of Fig.\ 3 shows that for $\sigma/J = 0.2$ when $T/J \lesssim 0.04$, $\langle E_{\textrm{B}} \rangle \lesssim -2.08J$, which we observe from Fig.\ 1 is the center of the LGS density of state for this disorder. As will be shown in Section \ref{Se:3.2}, sub-diffusive dynamics occurs when the temperature is low enough such that the  particle's thermal energy is so small that it resides deep within the LGS manifold of states.

\begin{figure}[tb]
\includegraphics[width=0.9\linewidth]{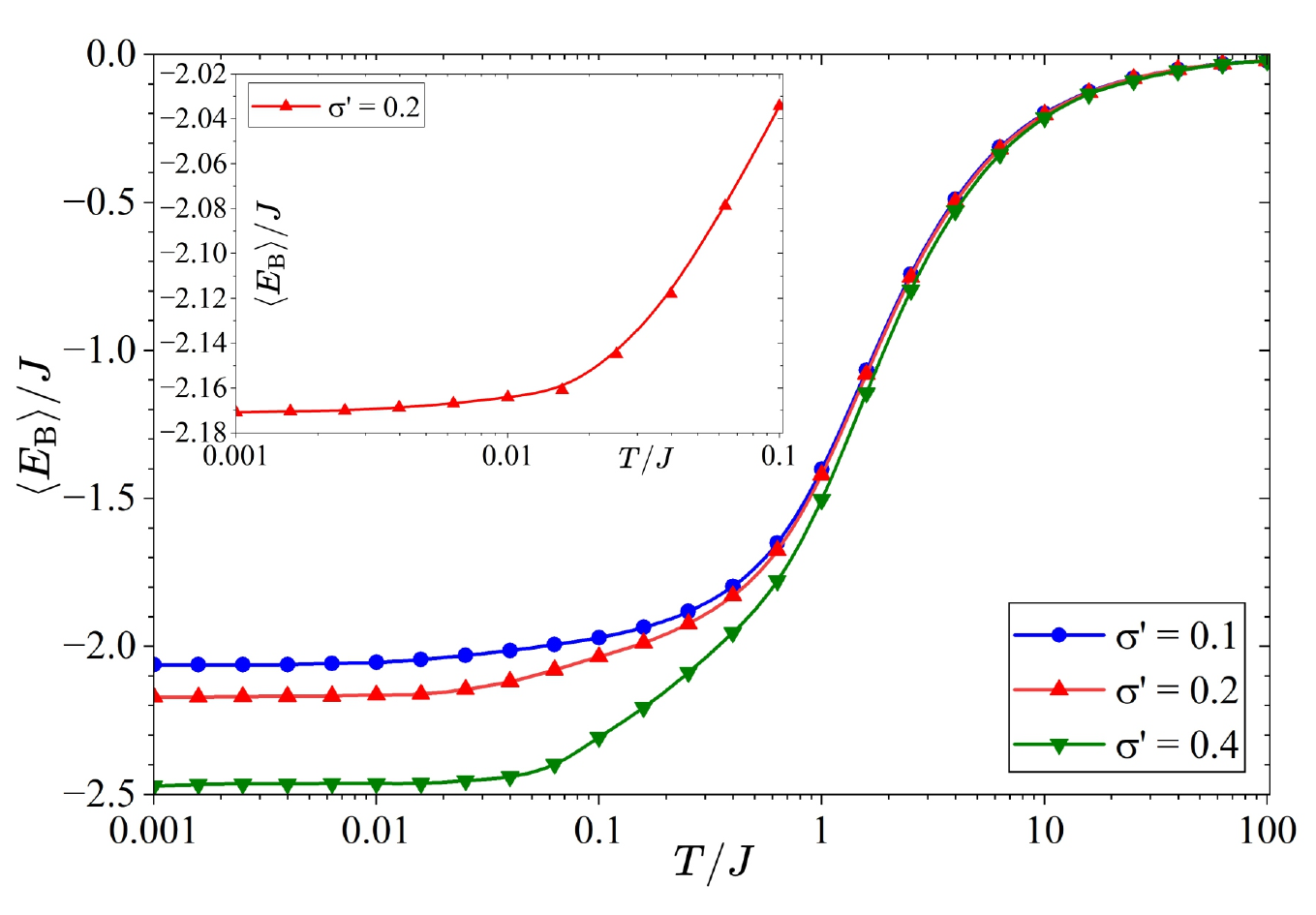}
\label{Fi:5}
\caption{Ensemble-averaged thermal energies, $\langle E_{\textrm{B}} \rangle$,  versus temperature ($k_B = 1$) for various disorder strengths, $\sigma' = \sigma/J$. Linear chains of 500 sites and 500 realizations of the disorder. }
\end{figure}

\subsection{Diffusion Coefficient in the High-Temperature Limit:  $T > J$}\label{Se:3.1}

The results described in  this Section using the velocity autocorrelation function (i.e.,  Eq.\ (\ref{Eq:15}))  were performed for chains of 400 sites and there were 200 realizations of the disorder, while results using  the eigenstate thermalization hypothesis used  chains of 400 sites and $10^5$ quantum jumps.

We first discuss the predictions of both of the methods described in Section \ref{Se:2.2} for translationally invariant systems. As illustrated by the black curves and symbols in Fig.\ 3, their predictions are identical. For $T \gg J$ they follow the HSR prediction that $D_{\infty} = 2J^2/\Gamma(T) = 2J^2/(\Gamma_0 \times T)$. This follows from Eq.\ (\ref{Eq:13}), because for $T \gg J$, $\langle V^2\rangle_{\textrm{th}} = 2J^2$. Conversely, for $T \ll J$, $D_{\infty} = 2JT/\Gamma(T) = 2T/\Gamma_0$, because now (from equipartition) $\langle V^2\rangle_{\textrm{th}} = 2JT$.\footnote{Returning to dimensionful units and using $J = \hbar^2/2md^2$ gives $\VV = k_BT/m$.} The crossover occurs at $T \sim J$.

The low-temperature result can be derived independently from Eq.\ (\ref{Eq:13}) by a simple  classical theory of scattering with a frequency-independent scattering rate (i.e., Brownian motion). The diffusion coefficient, $D = (N(t)/t) \ell^2/2$, where $N(t)$ is the number of scattering events in a time $t$, i.e., $N(t)/t = \Gamma$. $\ell$ is the mean-free-path, which satisfies $\ell^2 = \langle V^2\rangle \tau^2$, where $\tau^2 = 2/\Gamma^2$ is the mean-squared scattering time\cite{Barford2024}. Thus, $D = \langle V^2\rangle/\Gamma$, where $\langle V^2\rangle = 2JT$, and therefore $D = 2T/\Gamma_0$.

\begin{figure}[tb]
\includegraphics[width=0.9\linewidth]{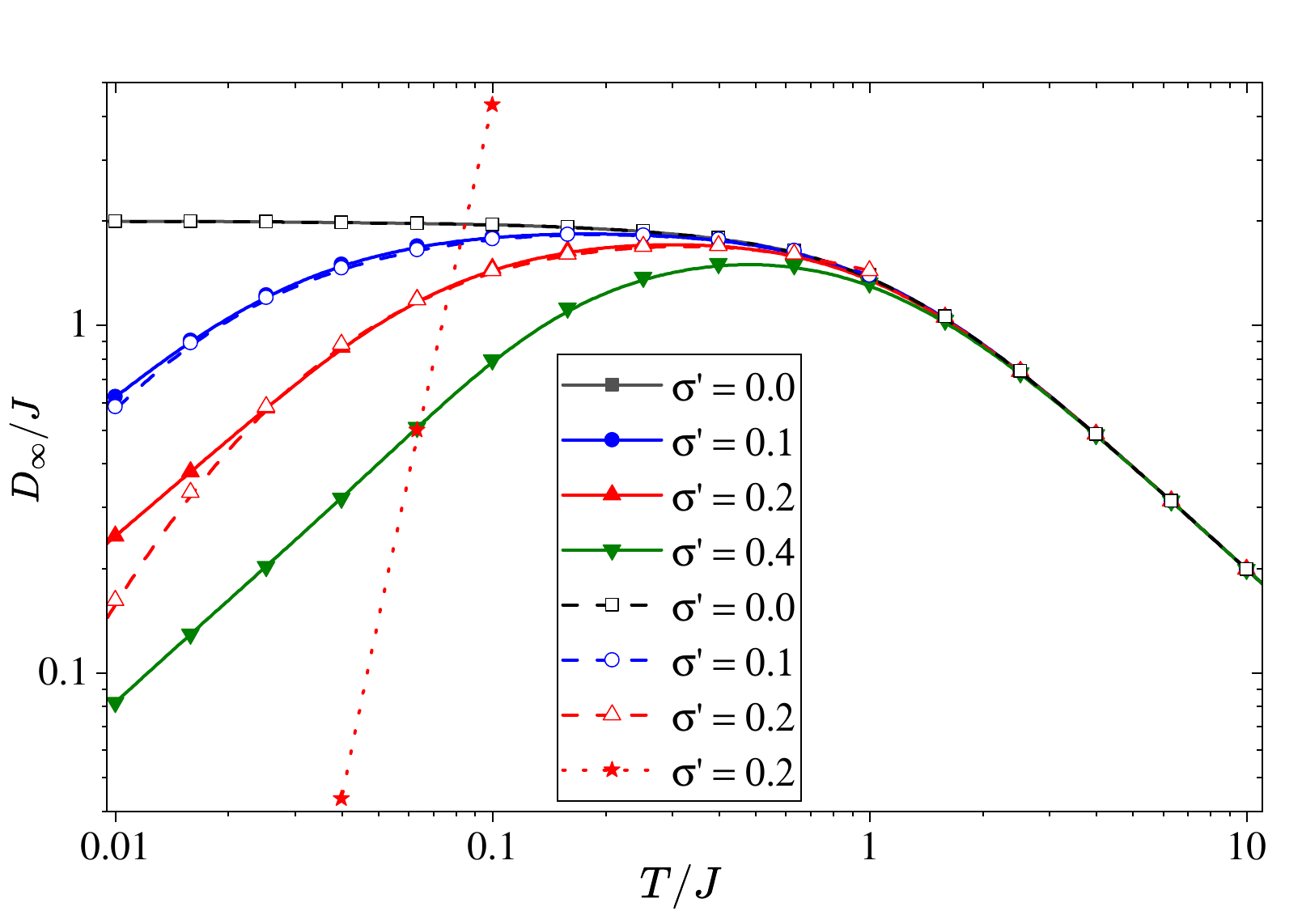}
\label{Fi:2}
\caption{The diffusion coefficient as a function of temperature for different values of the disorder, $\sigma' = \sigma/J$. The dephasing factor, $\Gamma_0 = 1$ (where $\Gamma(T) = \Gamma_0\times T$). ($k_B  = \hbar = d = 1$.)
Solid curves and filled symbols: results using the velocity autocorrelation function (i.e.,  Eq.\ (\ref{Eq:15})) with chains of 400 sites and 200 realizations of the disorder.
Dashed curves and open symbols: results from quantum trajectories and the  ETH with chains of 400 sites and $10^5$ quantum jumps.
The dotted curve with star symbols is the diffusion coefficient computed via the Redfield equation, as described in Sections \ref{Se:2.3} and \ref{Se:3.2}.
}
\end{figure}

We now turn to consider the effects of static disorder, where Anderson localization plays a role. This has been thoroughly investigated in the `high-temperature' HSR limit\cite{Cao2013,Knoester2021}, where the energy eigenstates are equally populated.  In this limit there are two regimes. At low dephasing rates the phenomenon of Environment Assisted Quantum Transport (EAQNT) occurs. This arises if the dephasing time is longer than the time taken for a wavefunction to become Anderson localized. The Anderson localization time is $\tl \sim \ell_{\textrm{loc}}/V$, where $\ell_{\textrm{loc}}$ is the Anderson localization length and  $V$
is the speed of a coherently dispersing wavepacket.
EAQNT occurs when $\tl  < \Gamma^{-1}$. Dephasing causes transport in this regime, because the interaction with the environment causes the wavefunction to stochastically collapse (as described in Section \ref{Se:2.2.2}). As the location of the collapse is random within the spatial span of the wavefunction, each collapse causes a root-mean-squared jump of size $\sim \el$. The diffusion coefficient is therefore $D_{\textrm{EAQNT}} \sim (N(t)/t)\el^2 = \Gamma(T) \el^2 \propto T$, as $\Gamma(T) = \Gamma_0 \times T$ and  $\el$ is independent of temperature in the high-T limit.

\begin{figure}[tb]
\includegraphics[width=0.9\linewidth]{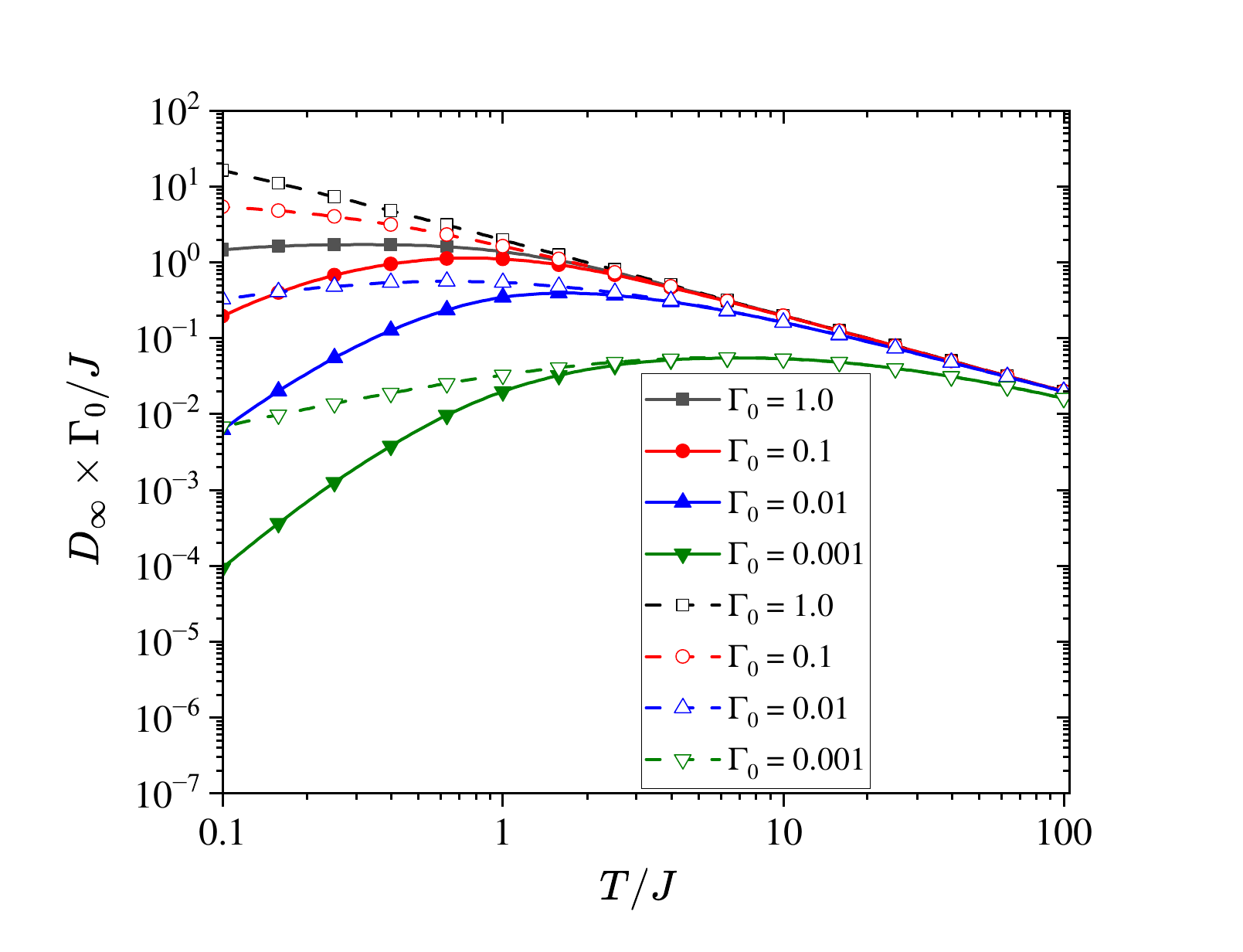}
\label{Fi:3}
\caption{Solid curves and filled symbols: the diffusion coefficient (via Eq.\ (\ref{Eq:15})) as a function of temperature for different values of the dephasing factor $\Gamma_0$ (where $\Gamma(T) = \Gamma_0\times T$)
when Boltzmann populations are enforced at finite temperatures, i.e., $P_a =P_a^{\textrm{B}}(T)$. The onsite disorder  $\sigma/J = 0.2$. ($k_B  = \hbar = d = 1$.)
Dashed curves and open symbols: the diffusion coefficient determined via  Eq.\ (\ref{Eq:15}) in the `high-T' limit when the eigenstates are equally populated, i.e., $P_a= 1/N$.
(Note that since $D_{\infty}$ is multiplied by $\Gamma_0$ the low temperature results  differ by a factor of $\Gamma_0^2$, whereas  the high temperature results converge to the same value.)}
\end{figure}

Conversely, at high dephasing rates the quantum-Zeno regime applies. Here, dephasing is too fast for the particle to become Anderson localized. Thus, the  root-mean-squared jump size is the mean-free path, $\ell = V/\Gamma$ and therefore $D_{\textrm{QZ}} \sim (N(t)/t)\ell^2 = V^2/\Gamma \propto T^{-1}$, as $V$ is independent of temperature in the high-T limit.
The crossover between the two regimes occurs when $D_{\textrm{EAQNT}} \sim D_{\textrm{QZ}}$, i.e., $\Gamma_c \sim V/\ell_{\textrm{loc}}$. These predictions of the high-T limit are confirmed in Fig.\ 5 for different dephasing parameters, $\Gamma_0$, by the dashed curves and open symbols: for a given disorder the cross-over temperature is, $T_c \sim \Gamma_0^{-1}$, while $D_{\infty}(T) \sim T$ for $T < T_c$ and $D_{\infty}(T) \sim T^{-1}$ for $T > T_c$.

The previous discussion applies to the HSR (high-T) limit when all eigenstates are equally populated, so that $V = \sqrt{2}J$ and $\ell_{loc} \sim \sigma^{-2}$. We expect there to be a qualitatively similar EAQNT to quantum-Zeno crossover when Boltzmann populations are enforced at finite temperatures. This is confirmed by Fig.\ 4, which shows $D_{\infty}(T)$  calculated via Eq.\ (\ref{Eq:15}) for different values of disorder for the case $\Gamma = T$ (i.e., $\Gamma_0 = 1$).
In all cases there is a low temperature EQANT regime where $D_{\infty}(T)$ is an increasing function of $T$, and a cross-over to a higher temperature quantum-Zeno regime where $D_{\infty}(T)$ follows the disorder-free behavior (denoted by the black curve) and decreases with $T$. Also shown in Fig.\ 4  are some predictions using quantum trajectories and the  eigenstate thermalization hypothesis. Despite the differing assumptions of the two methods, except at the lowest temperatures there is good agreement with the predictions derived from Eq.\ (\ref{Eq:15}).

Fig.\ 5 also shows $D_{\infty}(T)$ computed via Eq.\ (\ref{Eq:15}) when Boltzmann populations are enforced at finite temperatures for different dephasing parameters, $\Gamma_0$. As $\Gamma_0$ increases the EAQNT to QZ transition occurs at progressively lower temperatures and thus the deviations from the high-T limit of equal eigenstate populations increases. In particular, since both $V$ and $\el$ are temperature dependent for $W_{\textrm{LGS}} < T < J$, $\Gamma_c \sim V/\ell_{\textrm{loc}}$ has a complex temperature-dependency. We also note that in the EAQNT regime when $T < J$ the predicted diffusion coefficient is significantly reduced from the `high-T' values when thermal populations are imposed because of the temperature-dependency of $\el$ caused by the thermal excitations of QES.

For temperatures significantly smaller than the particle bandwidth the two assumptions of classical white-noise and that particle transport is determined by localization caused by dynamical disorder are no longer valid. Instead, the bath autocorrelation function becomes non-Markovian and transport is determined by hopping between Anderson-localized eigenstates. In addition, for temperatures smaller than the LGS bandwidth, the transport becomes non-diffusive. This is regime is described in the next section.

\subsection{Non-diffusive Dynamics when $T <  W_{\textrm{LGS}}$}\label{Se:3.2}

Sub-diffusive dynamics occurs for systems with rugged energy landscapes and is thus expected for temperatures much less than the exciton bandwidth, when exciton migration occurs across this landscape\cite{Vlaming2013}.
It has been observed experimentally in organic molecules\cite{Akselrod2014} and molecular fibers\cite{Hildner2024}, and predicted theoretically in molecular fibers\cite{Hildner2024} and conjugated polymers\cite{Perez}.

The MSD for subdiffusive processes satisfies,
\begin{equation}\label{Eq:39}
 \textrm{ MSD}(t) = 2 D(\alpha) t^{\alpha}
\end{equation}
with $\alpha < 1$.
Here it is computed via
\begin{equation}\label{}
  \textrm{MSD}(t) = \textrm{Var}(t) - \textrm{Var}(0),
\end{equation}
where Var($t$) is the variance of the particle-density distribution defined by
\begin{equation}\label{Eq:60}
  \textrm{Var}(t)  = \sum_n n^2 P_n(t) - \left(\sum_n n P_n(t)\right)^2,
\end{equation}
and $P_n(t) = \sum_a |\psi_{na}|^2 P_a(t)$ is the particle-density on site $n$ at time $t$. The eigenstate population, $P_a(t)$, is determined by Eq.\ (\ref{Eq:50}).

An example of an ensemble-averaged MSD versus $t$ is illustrated in Fig.\ 6 for a linear chain of 800 sites with $\sigma/J = 0.2$ and $T/J = 0.0158$. We observe three time regimes. Initially, as shown in blue, the MSD increases linearly in time. For these short times the average MSD is smaller than one repeat unit, so the particle is not sampling the energetically disordered landscape. Second, as shown in red, for time spanning ca.\ five orders of magnitude the average MSD increases sub-linearly. Finally, the average MSD saturates to its thermal equilibrium given by Eq.\ (\ref{Eq:60}) with
  $P_n(t) \rightarrow P_n^{\textrm{B}} = \sum_a |\psi_{na}|^2 P_a^{\textrm{B}}$.

As the particle's wavepacket evolves at  a fixed temperature from a single energy eigenstate it spreads over an increasing number of  eigenstates. During this evolution its energy fluctuates, as illustrated in  the inset of Fig.\ 6 which shows the root-mean-squared fluctuation in the energy, $\langle \Delta E \rangle$, as a function of time. Throughout the  evolution $\langle \Delta E \rangle < T/2$ and it becomes negligible at equilibration, because the heat capacity $C \rightarrow 0$ for $T < \Wlgs$.

\begin{figure}[tb]
\includegraphics[width=0.9\linewidth]{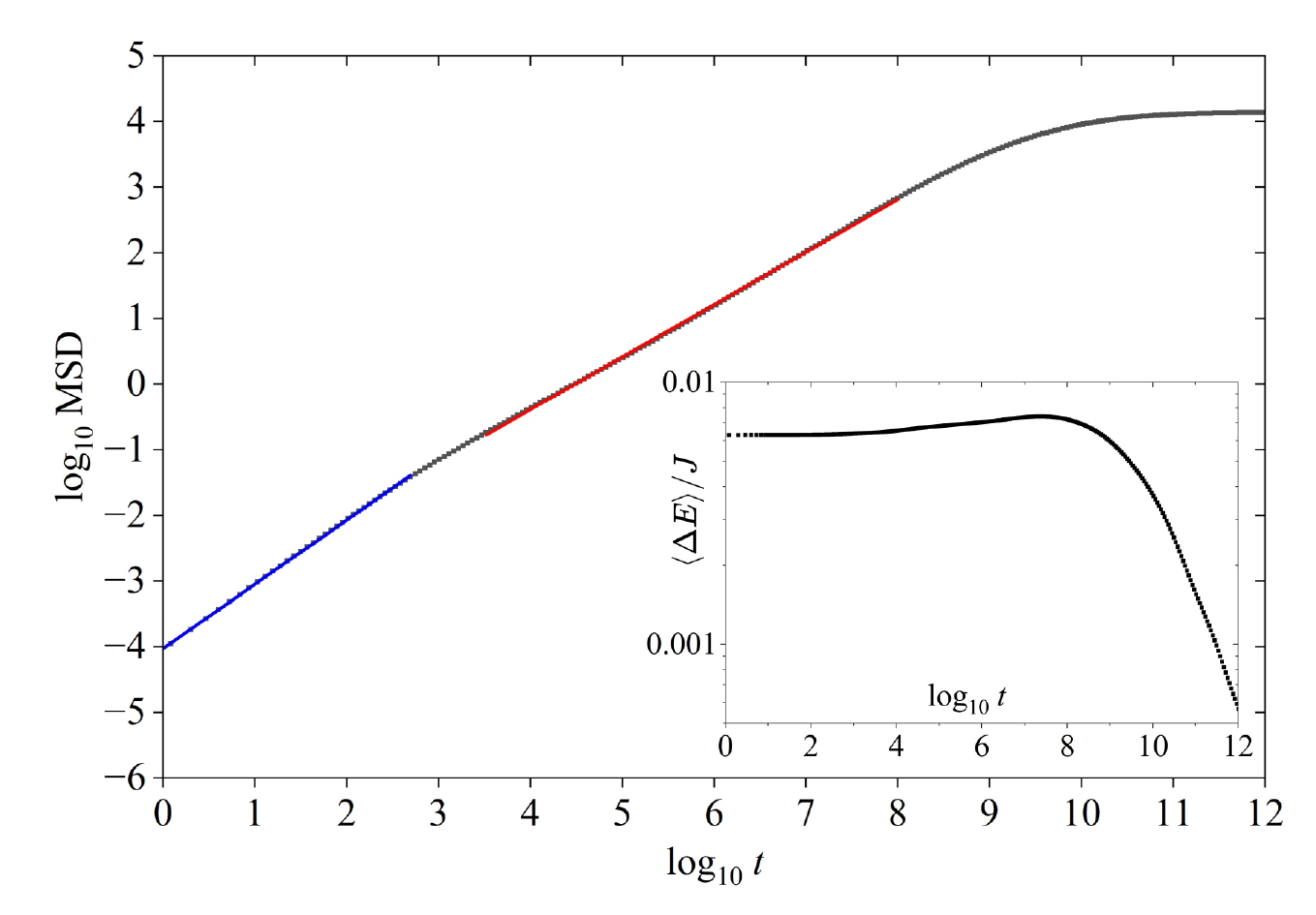}
\label{Fi:6}
\caption{$\log_{10} \textrm{MSD}$ versus $\log_{10}t$. Results for $\sigma/J = 0.2$, $T/J = 0.0158$, 800 sites and 1000 realizations of the disorder.
The blue and red parts of the curve are linear fits. The blue part indicates transient diffusive dynamics, while
the red part indicates the regime of sub-diffusive dynamics.
The inset shows the root-mean-squared fluctuation in the energy,  $\langle \Delta E \rangle$.
Time in units of $J^{-1}$ with $k_B  = \hbar = d = 1$.}
\end{figure}

Fig.\ 7 shows a finite-size extrapolation of $\alpha$ to infinite chain lengths for $\sigma/J = 0.2$ and two values of $T$ in the sub-diffusive regime. Evidently, $\alpha(N \rightarrow \infty) < 1$ and $\alpha$ increases with increasing temperature.
Ideally, one would like to perform such finite-size extrapolations for a variety of disorder and temperatures, but unfortunately this was generally unreliable because of the numerical inaccuracies with the matrix diagonalization of the rate matrix, $\textbf{K}$, as described in Section \ref{Se:2.3}. Accordingly, $\alpha$  is plotted as a function of temperature in Fig.\ 8  for various values of the static disorder  for chains of 800 sites.

\begin{figure}[tb]
\includegraphics[width=0.9\linewidth]{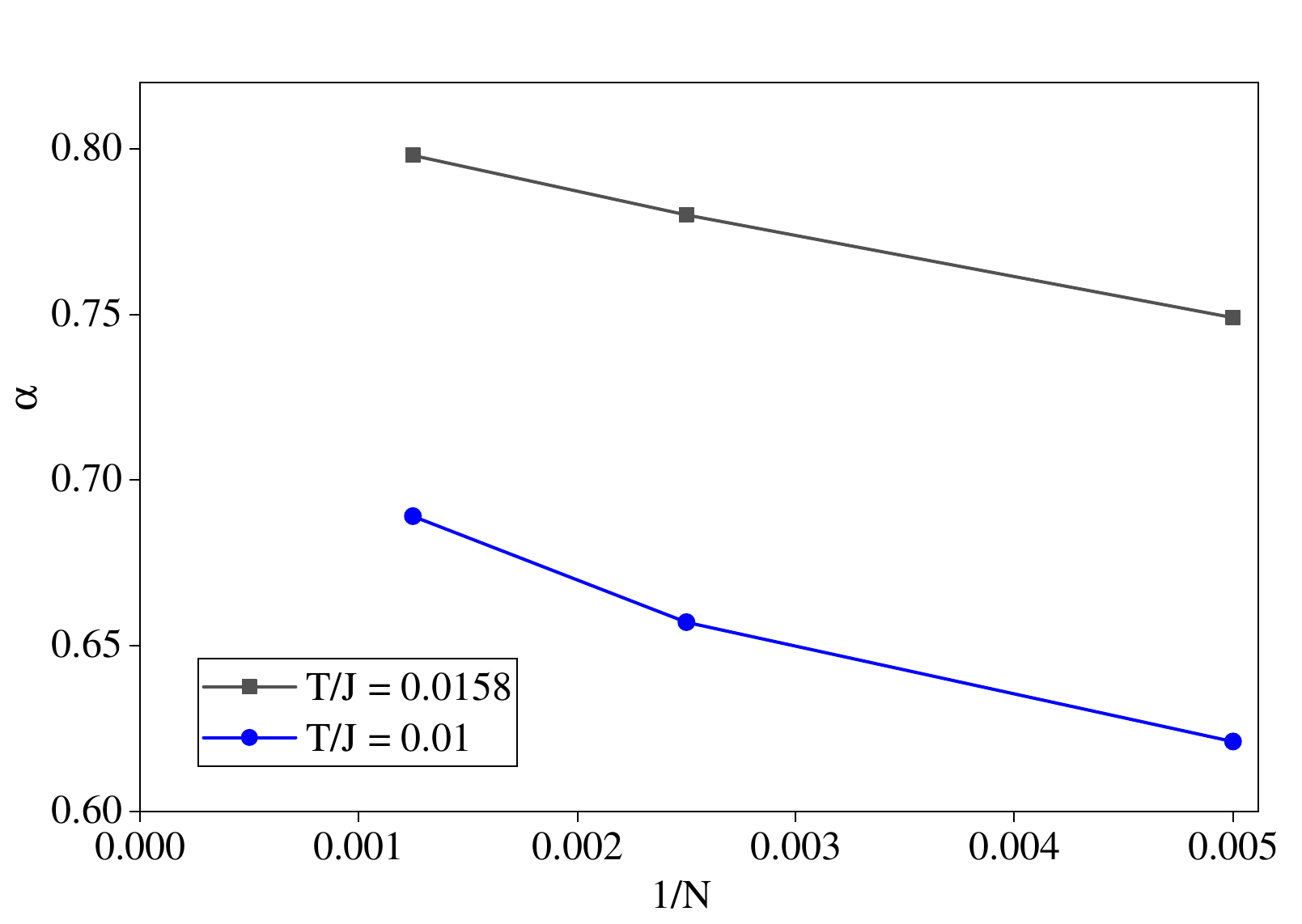}
\label{Fi:7}
\caption{
$\alpha$ versus $1/N$ (where $N$ is the number of sites) for $\sigma/J = 0.2$.
$\alpha(1/N \rightarrow 0) \rightarrow 0.814$ and $0.707$ for $T/J = 0.0158$ and $0.01$, respectively.
$k_B = 1$.
}
\end{figure}

\begin{figure}[tb]
\includegraphics[width=0.9\linewidth]{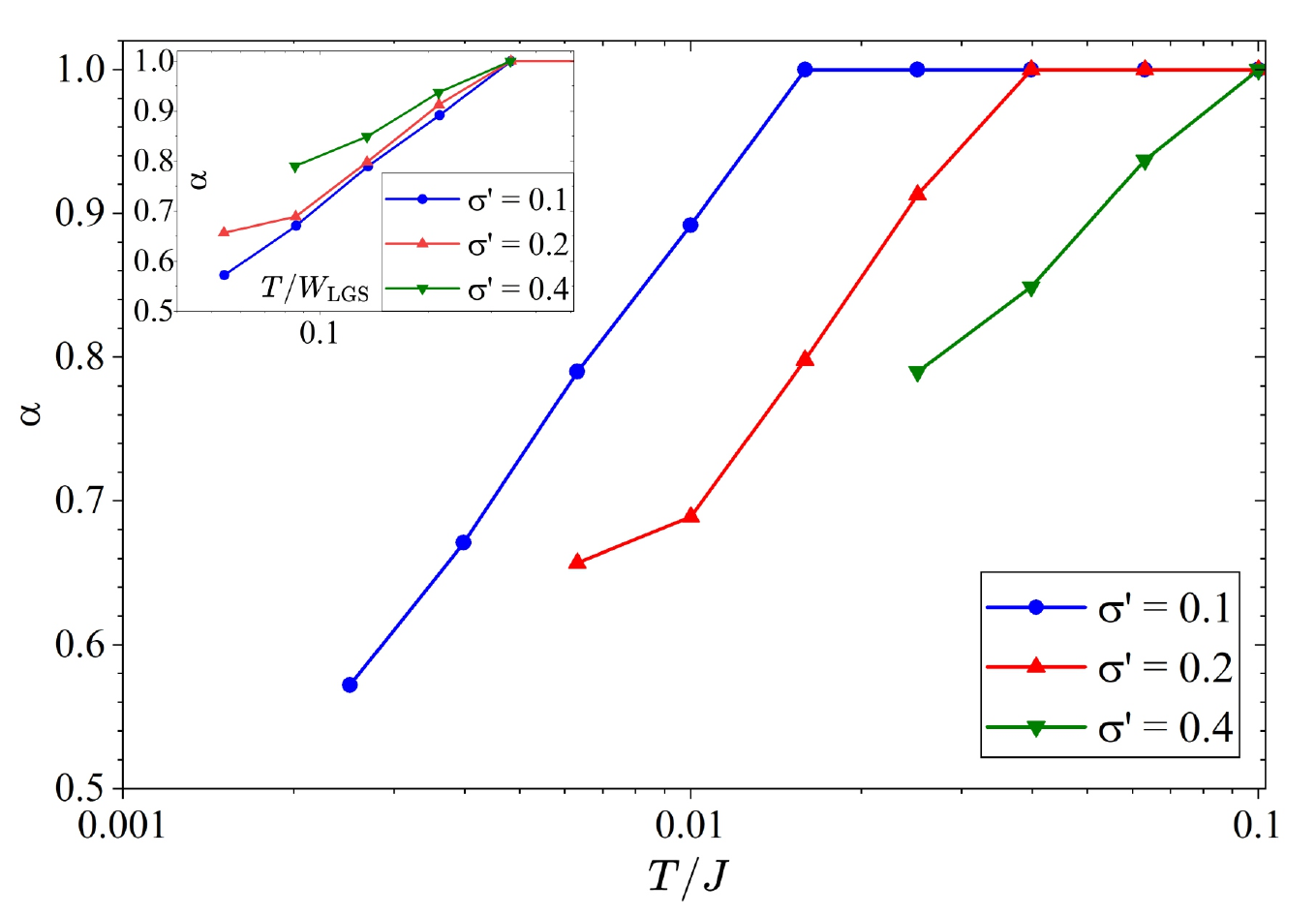}
\label{Fi:5}
\caption{
$\alpha$ versus $T$ for different disorder values, $\sigma' =\sigma/J$ and chains of 800 sites. The inset shows $\alpha$ versus the scaled temperature, $T/W_{\textrm{LGS}}$, where $W_{\textrm{LGS}} \sim  J(\sigma/J)^{4/3}$ is the width of the LGS density of states (shown in Fig.\ 1). $k_B = 1$.
}
\end{figure}

The exponent $\alpha$ is a decreasing function of disorder. In all cases $\alpha \rightarrow 1$ as $T/J \rightarrow 0.1$, indicating that for temperatures greater than the LGS bandwidth, $W_{\textrm{LGS}}$, the transport becomes diffusive.
Since for exciton migration in the limit that $T  < W_{\textrm{LGS}} \sim J(\sigma/J)^{4/3}$, the only energy scales are $T$ and $W_{\textrm{LGS}}$, we speculate that $\alpha$ is a function of the dimensionless ratio $T/W_{\textrm{LGS}}$. Thus, the inset of Fig.\ 8 shows $\alpha$ versus $T/W_{\textrm{LGS}}$ for three different disorder values. Although the curves do not exactly collapse onto the same curve, given that a finite-size scaling analysis has not been possible, the scaling assumption seems reasonable. In addition, we observe from the inset of Fig.\ 8 that $\alpha \rightarrow 1$ as $T/W_{\textrm{LGS}} \rightarrow 1$.

We now examine in more detail the cross-over from subdffusive to diffusive dynamics when   $\sigma/J = 0.2$. As shown in Fig.\ 8, this crossover occurs at $T/J \sim 0.04$. Returning to the discussion at the start of this Section, Fig.\ 1 and the inset of Fig.\ 3 show that for $T/J \le 0.04$, the mean thermal energy, $\langle E_{\textrm{B}} \rangle$, lies deep within the LGS manifold of states, and thus subdiffusive behavior occurs when only LGS participate in the dynamics.

\begin{figure}[tb]
\includegraphics[width=0.9\linewidth]{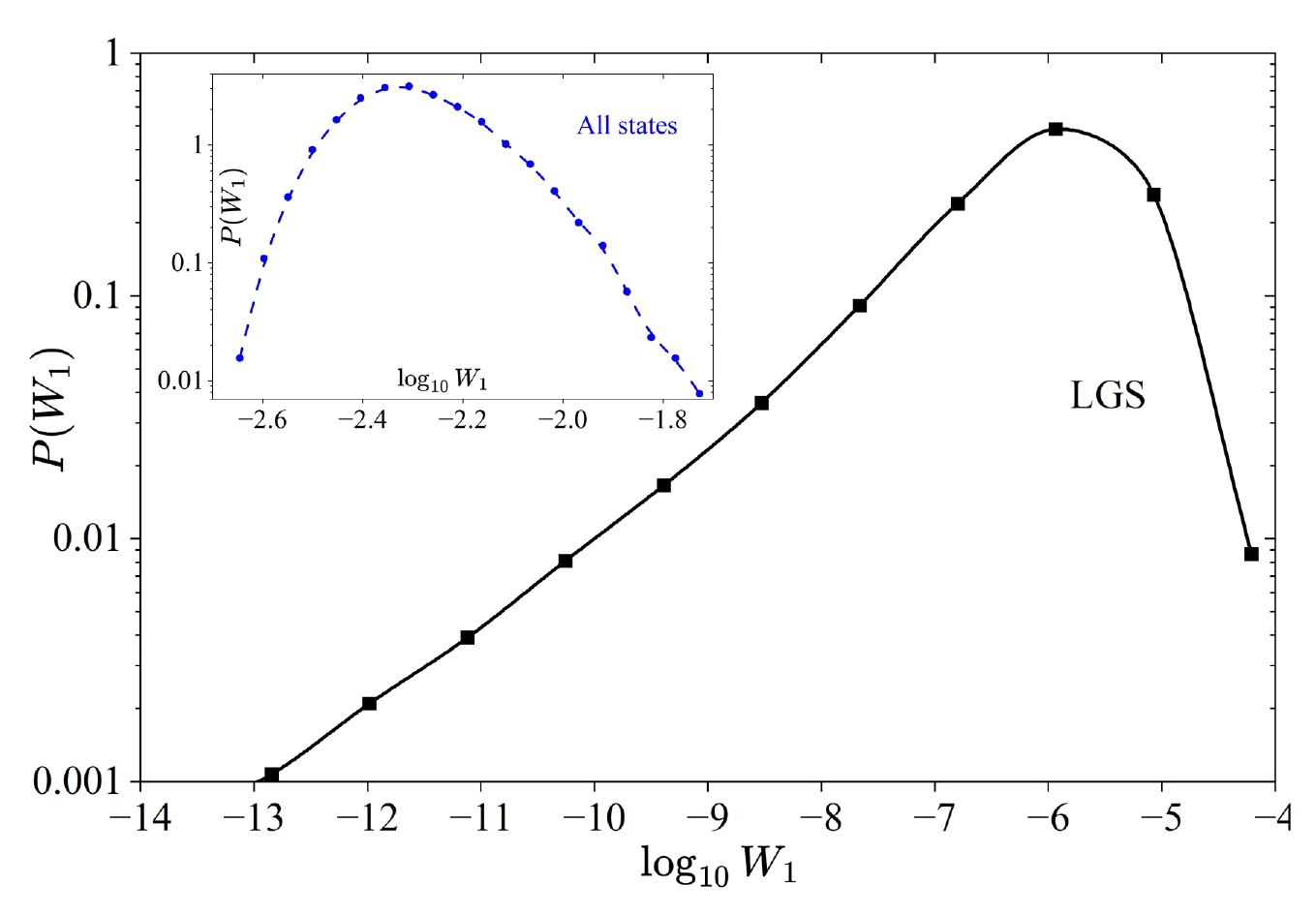}
\label{Fi:5}
\caption{The distribution of transition rates between neighboring sites, defined by Eq.\ (\ref{Eq:700}) and Eq.\ (\ref{Eq:701}). The main figure shows $P(W_1)$ evaluated with only LGS, while the inset shows $P(W_1)$ when all states are included. $T/J = 0.0158$ and $\sigma/J = 0.2$. }
\end{figure}

Sub-diffusive dynamics in linear systems described by Eq.\ (\ref{Eq:1}) has been investigated by Eisfeld \emph{et al.}\cite{Eisfeld2010,Vlaming2013} for the case where the site energies satisfy a heavy-tailed L\'evy distribution. According to ref\cite{Vlaming2013}, the  quantity that helps explain sub-diffusive behavior is the probability distribution for the average transition rate between neighboring sites, defined as
\begin{equation}\label{Eq:700}
  W_1 = \frac{1}{N-1}\sum_{n=1}^{N-1} \tilde{W}_{n,n+1},
\end{equation}
where
\begin{equation}\label{Eq:701}
\tilde{W}_{mn} = \sum_{ab} \bar{k}_{ab} |\psi_{ma}|^2 |\psi_{nb}|^2
\end{equation}
and $\bar{k}_{ab}=(k_{ab}+k_{ba})/2$.
A (non-Gaussian) L\'evy distribution in site energies results in a bimodal  $P(W_1)$, with $P(W_1) \rightarrow \textrm{finite}$ as $W_1 \rightarrow 0$. This distribution for transition rates represents outliers for  transitions from long-lived trapping sites.

Fig.\ 9 shows $P(W_1)$ for a Gaussian distribution of site energies. The inset of Fig.\ 9 is $P(W_1)$ computed from Eq.\ (\ref{Eq:700}) when all energy eigenstates are included in the sum over $a$ and $b$ in Eq.\ (\ref{Eq:701}).
This is valid for higher temperatures when diffusive dynamics occurs entirely within the manifold of all states, as indicated by the density of states in Fig.\ 1.
In this case, $P(W_1)$ shows a narrow single peak with a tail towards large values of $W_1$. The main figure, however, shows $P(W_1)$ computed from Eq.\ (\ref{Eq:701}) when only LGS are included in the sum. This is valid for the low temperatures when sub-diffusive dynamics occurs entirely within the manifold of LGS. Again, there is a single peak, but now it is very wide and  strongly skewed to very small values of $W_1$.

\begin{figure}[tb]
\includegraphics[width=0.9\linewidth]{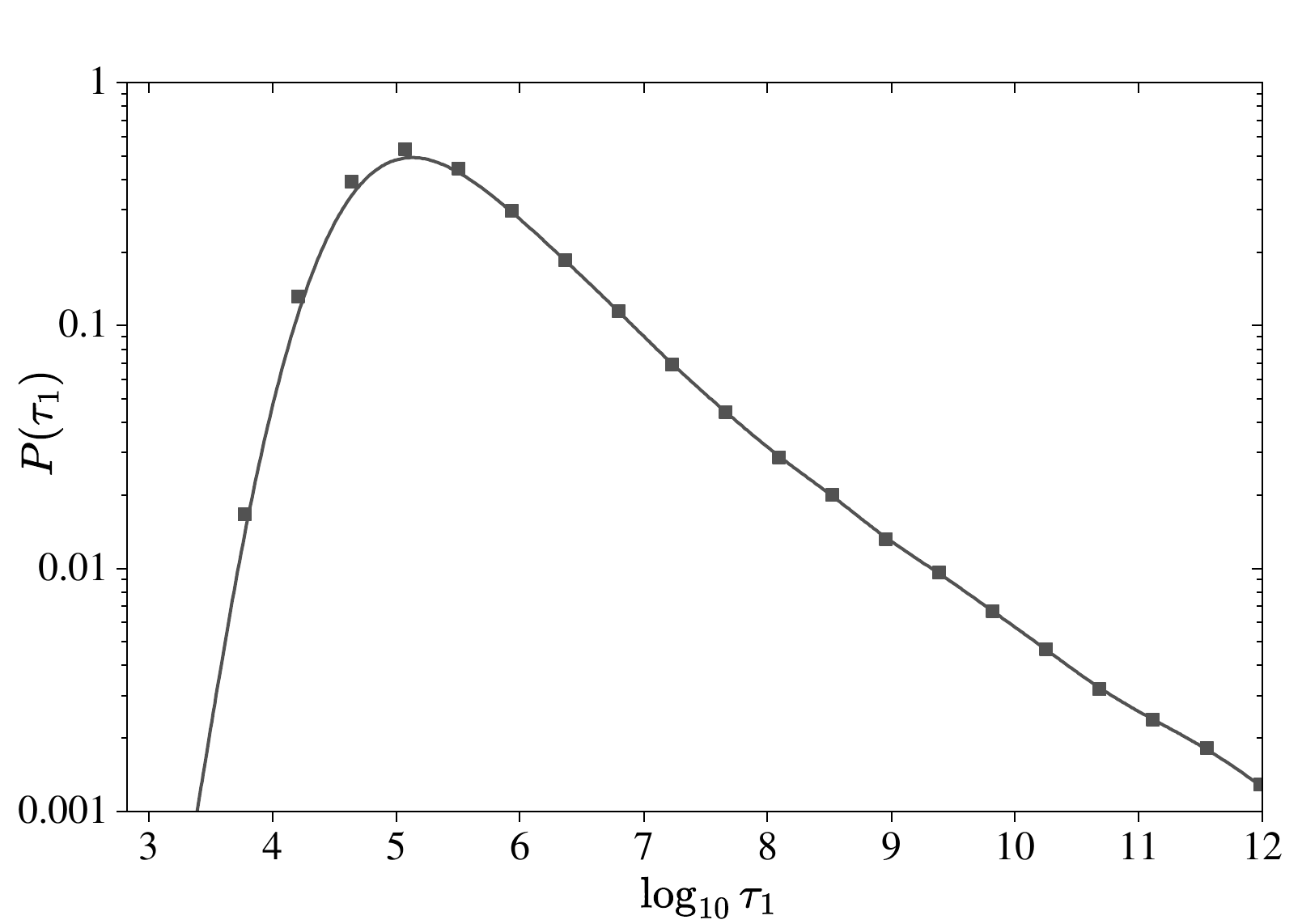}
\label{Fi:5}
\caption{The distribution of transition times between neighboring sites,  $P(\tau_1)$, evaluated with only LGS. $T/J = 0.0158$ and $\sigma/J = 0.2$.
For $\tau_1 > 10^5$ (in units of $\hbar/J$), $P(\tau_1) \sim \tau_1^{-\gamma}$, where $\gamma \sim 0.39$. }
\end{figure}

If $W_1$ is the average transition rate between neighboring sites, we may interpret its inverse as the average transition time between sites, i.e., $\tau_1 = W_1^{-1}$. The probability distribution of  $P(\tau_1)$ when only LGS contribute to Eq.\ (\ref{Eq:701}) is displayed in Fig.\ 10. This shows  that for $\tau_1 > 10^5$ s (in units of $\hbar/J$), $P(\tau_1) \sim \tau_1^{-\gamma}$, where $\gamma \sim 0.39$. Interestingly, as shown in Fig.\ 6, a timescale of $10^5$ s is approximately the same timescale at which (for this disorder value) subdiffusive dynamics begins. Thus, we conclude, as also observed by Carta \emph{et al.}\cite{Hildner2024}, a non-Gaussian L\'evy distribution of the site energies is not a necessary condition for subdiffusive dynamics.

Finally, we turn to discuss the diffusive dynamics in the low-temperature regime. Fig.\ 8 indicates that when $\sigma/J = 0.2$ the particle motion becomes diffusive for $T/J \gtrsim 0.04$. The diffusion coefficient is therefore  computed in this diffusive regime using  Redfield theory and plotted in Fig.\ 4. The quantitative agreement with the values computed via the velocity autocorrelation function (i.e., Eq.\ (\ref{Eq:15})) are poor, indicating both a failure of Redfield theory at higher temperatures and  larger coupling, and a failure of the velocity autocorrelation function method at lower temperatures when thermal populations are enforced and the classical bath approximation breaks down.

\section{Conclusions}\label{Se:4}

This paper has presented calculations of  temperature-dependent transport for  charges and excitons in  one-dimensional
systems subject to static and dynamic disorder.
The transport properties have been determined by three complementary methods. One approach is via the time-integration of the velocity autocorrelation function  in the secular limit, i.e., when the energy eigenstate coherences are neglected. The second approach is via the mean-squared-displacement of thermal wavepackets subject to stochastic collapse via Lindblad jump operators. These two methods are applicable in the high-temperature regime, where the noise is temporally uncorrelated. In this regime the noise causes particle localization and the transport is diffusive. The third approach -- applicable in the low-temperature regime -- is  weak-coupling Redfield theory. Here, static disorder causes particle localization. The dynamics is non-diffusive for thermal energies deep within the manifold of local-ground-states (LGS).

The general conclusions of this paper are the following:
\begin{enumerate}
\item{For a \emph{uniform} system and a classical bath, when $J/N \ll T < J$,  $\Dinf = 2J/\Gamma_0$, i.e., independent of temperature; this corresponds to the motion of a classical Brownian particle. Conversely, when $T > J$, $\Dinf = 2J^2/(\Gamma_0 \times T)$; this is the `high-T' HSR limit, when all the energy eigenstates are populated. The cross-over occurs when $T \sim J$.}
\item{For  \emph{disordered} systems:
\begin{enumerate}
\item{For a fixed temperature and dephasing factor, $\Gamma_0$, $\Dinf$ decreases as the static disorder, $\sigma$, increases.}
\item{For a fixed temperature and  static disorder, $\sigma$, $\Dinf$ decreases as the dephasing factor, $\Gamma_0$, increases.}
\item{For any value of  $\Gamma_0$ and $\sigma$,  $\Dinf$ is a non-monotonic function of $T$, increasing  for $T < T_c$ and decreasing for $T > T_c$ when it follows the disorder-free behavior.}
\item{$T_c(\sigma,\Gamma_0)$ is a complex function of $\sigma$ and $\Gamma_0$, but broadly, $T_c$ increases as $\Gamma_0$ increases for fixed $\sigma$ or as $\sigma$ increases for fixed $\Gamma_0$.}
\end{enumerate}}
\item{For a fixed temperature and for any value of $\Gamma_0$ and $\sigma$, $\Dinf(T)$ is smaller than  $\Dinf$ evaluated in the `high-T' limit (i.e., when all eigenstates are equally populated).}
\item{For temperatures smaller than the LGS bandwidth there is a regime of sub-diffusive dynamics where $\textrm{MSD} = 2Dt^{\alpha}$, with $\alpha < 1$. Here, $\alpha$ is a decreasing function of disorder and an increasing function of temperature, such that $\alpha \rightarrow 1$ when $T/\Wlgs \sim 1$. As indicated in Fig.\ 8, $\alpha$ appears to be a universal function of the dimensionless ratio $T/\Wlgs$.}
\end{enumerate}

There is obvious scope for extensions of this work, e.g., describing particle dynamics in a uniform system using a quantum system-bath autocorrelation function, incorporating correlated static disorder\cite{Hildner2024}, and the inclusion of explicit electron-phonon interactions.




%

\end{document}